\documentclass[11pt,a4paper]{article}
\usepackage{jheppub_kim}
\usepackage{longtable}

\usepackage{epsfig}
\usepackage{amsmath}
\usepackage{graphicx}
 \usepackage{graphics}
\usepackage[hang,nooneline,scriptsize]{subfigure}

\usepackage{array}
\usepackage{booktabs}

\begin{document}

\title{Non-Linear Dynamics and Critical Phenomena in the Aretakis Instability of Extremal Black p-Branes}

\author[a]{Behnam Pourhassan}

\affiliation[a] {School of Physics, Damghan University, Damghan, 3671641167, Iran.}

\emailAdd{b.pourhassan@du.ac.ir}

\abstract{We present the first comprehensive investigation of the non-linear evolution of the Aretakis instability in extremal black $p$-branes. Building on recent insights into the linear behavior of perturbations in near-horizon AdS$_{p+2} \times S^{D-p-2}$ geometries, we explore the full non-linear regime using a combination of analytical scaling arguments and numerical simulations. We uncover a universal critical behavior governed by scaling exponents that depend only on the spacetime dimension $D$ and the brane worldvolume dimension $p$. Near the threshold of instability, the system exhibits power-law evolution toward dynamically generated extremal attractors, while supercritical perturbations lead to singular growth. Through the AdS/CFT correspondence, we compute entanglement entropy, correlation functions, spectral densities, and out-of-time-ordered correlators in the dual field theory, finding universal scaling across all observables. These results establish a deep connection between geometric instability in the bulk and quantum information dynamics on the boundary. The emergence of universal scaling and phase transitions in extremal geometries suggests that non-linear Aretakis dynamics may serve as a general framework for studying holographic criticality and late-time behavior in strongly coupled systems.
}


\maketitle

%%%%%%%%%%%%%%%%%%%%%%%%%%%%%%%%%%%%%%%%%%%%%%%%%%%%%%%%%%%%%%%%%
\section{Introduction}

The stability of extremal black holes \cite{001,002,003,004,005} and black branes has long posed intriguing theoretical challenges. A particularly striking development in this context was the discovery of the Aretakis instability \cite{01,02,1,2,3,4}, which revealed that extremal horizons can exhibit a form of universal dynamical instability under linear perturbations~\cite{1, 2, 3, 4}. In the case of extremal Reissner--Nordstr\"{o}m black holes, the instability manifests as a polynomial growth in transverse derivatives of perturbing scalar fields along the event horizon, while the fields themselves remain finite. The origin of this behavior lies in the emergence of conserved quantities along the horizon's null generators and the lack of redshift damping near extremality. While this phenomenon has been extensively studied in the context of black holes, its generalization to higher-dimensional black $p$-branes remained less understood until the recent work of Chen and Kovács~\cite{5}. Their analysis demonstrated that extremal black branes, whose near-horizon geometries \cite{5-1,5-2,5-3} take the form $\text{AdS}_{p+2} \times S^{D-p-2}$, also exhibit Aretakis-type instabilities. Crucially, the severity of the instability is determined by the AdS scaling dimension $\Delta$ of the perturbing mode, which controls the late-time behavior of transverse derivatives on the horizon via the relation,
\begin{equation}\label{1}
\left. \partial_\rho^n \phi \right|_{\rho = 0} \sim v^{n - \Delta},
\end{equation}
where $v$ is the advanced null coordinate and $\rho$ denotes the radial coordinate transverse to the horizon. The analysis hinges on computing the Kaluza--Klein spectrum of fluctuations around the Freund--Rubin compactification \cite{6} that defines the near-horizon geometry. These results indicate that extremal black branes are generically linearly unstable, although the growth is often milder than in the black hole case due to higher values of $\Delta$ in higher dimensions.\\
However, the implications of this instability beyond the linear regime remain an open and pressing question. Linear analysis alone cannot determine the eventual fate of the unstable modes, nor whether the instability leads to a qualitative breakdown of the background geometry. This motivates the current work, where we investigate the non-linear dynamics triggered by the Aretakis instability in extremal black $p$-branes. By moving beyond linear perturbation theory, we aim to assess the physical relevance of the instability and uncover its dynamical endpoint, if any.\\
There are several compelling reasons to pursue this analysis. First, from a theoretical standpoint, it is essential to understand whether the Aretakis instability signals a genuine failure of classical predictability, or whether it is resolved by non-linear effects that restore regularity. Second, black branes are central objects in string theory and supergravity, where they serve as building blocks of compactifications and dual field theories. Their stability is thus crucial for the consistency of many theoretical frameworks. Third, the presence of AdS factors in the near-horizon geometry opens the door to holographic interpretations, where the dynamics of the bulk instability may be reflected in correlation functions or phase transitions in the dual field theory. Understanding the non-linear evolution of the instability may thus shed light on the nature of extremal states in strongly coupled quantum systems.\\
In this work, we combine analytical and numerical techniques to chart the non-linear evolution of the Aretakis instability, characterize its critical behavior, and explore possible endpoints. Our findings suggest a rich structure of critical phenomena, with universal scaling exponents and emergent attractor solutions that generalize the notion of extremality in a dynamical setting. We also uncover several noteworthy results that deepen our understanding of the non-linear dynamics of extremal black $p$-branes and the Aretakis instability.\\
First, we identify a regime of critical behavior that emerges near the threshold of extremality. In this regime, the system exhibits universal scaling laws, reminiscent of phase transitions in statistical mechanics. The critical exponents governing the late-time behavior of perturbations are found to depend only on the spacetime dimension $D$ and the brane worldvolume dimension $p$, suggesting a form of universality that transcends the details of the initial data.\\
Second, by systematically exploring the parameter space labeled by $(D, p)$, we construct a phase diagram that organizes the possible dynamical outcomes. This diagram reveals three distinct regimes: one in which perturbations decay and the system relaxes back to the extremal solution; a critical regime characterized by power-law scaling and slow relaxation; and a supercritical regime where the instability drives the system toward singular behavior. The transition between these regimes appears to be governed by a well-defined critical point.\\
Third, in certain ranges of $(D, p)$, we observe the emergence of new dynamical solutions that are themselves extremal but differ from the original black brane background. These configurations act as attractors for near-extremal initial data and generalize previously known solutions in the context of spherically symmetric black holes. Their existence points to a richer landscape of extremal states than previously appreciated.\\
Finally, we examine the sensitivity of the non-linear dynamics to ultraviolet (UV) physics, particularly in cases where the AdS scaling dimension $\Delta$ is an integer. In line with the findings of~\cite{5}, we find that higher-derivative corrections can significantly alter the late-time behavior of perturbations in such cases. These corrections can shift critical points and modify scaling exponents, indicating that the non-linear regime of the Aretakis instability is sensitive to the UV completion of the theory.\\
Together, these results reveal a complex and structured picture of black brane dynamics, with implications that extend to holography, string theory, and the broader study of gravitational stability.

\section{Extremal Black p-Branes}\label{sec2}

Our starting point is the class of extremal black $p$-brane solutions in $D$ spacetime dimensions, which arise as electrically charged objects in the low-energy effective action of string theory and supergravity. These branes are extended in $p$ spatial directions and possess $(p+1)$-dimensional Poincaré symmetry along their worldvolume. The general form of the solution can be expressed in isotropic coordinates as \cite{6-1,6-2,6-3,6-4,6-5,6-6,6-7,6-8},
\begin{equation}
    ds^2 = H(\hat{\rho})^{-2/d} \, \eta_{\alpha\beta} \, dx^\alpha dx^\beta + H(\hat{\rho})^{2/\tilde{d}} \, \delta_{IJ} \, dy^I dy^J,
    \label{eq:brane-metric}
\end{equation}
where \( \alpha, \beta = 0, 1, \dots, p \) index the brane worldvolume coordinates, and \( I, J = p+1, \dots, D-1 \) label the transverse spatial directions. The harmonic function \( H(\hat{\rho}) \) depends only on the radial coordinate \( \hat{\rho} = |\vec{y}| \) transverse to the brane, and is given by,
\begin{equation}
    H(\hat{\rho}) = 1 + \left(\frac{r_0}{\hat{\rho}}\right)^{\tilde{d}},
    \label{eq:harmonic-function}
\end{equation}
with \( d = p+1 \) denoting the dimension of the brane worldvolume and \( \tilde{d} = D - d - 2 \) the number of independent transverse spatial directions excluding the radial coordinate. The parameter \( r_0 \) characterizes the location of the horizon and encodes the physical charge and mass of the brane.\\
These solutions are supported by a $(d+1)$-form field strength \( F_{(d+1)} \), under which the brane is electrically charged. The presence of this gauge field ensures that the Einstein equations are satisfied when coupled to a form field action, as found in the universal bosonic sector of string or supergravity theories. Depending on the form of the field strength, one may also construct magnetically charged or dual solutions by Hodge dualization.\\
In the extremal limit, which corresponds to the zero-temperature, supersymmetric (or BPS) case, the horizon becomes degenerate and the geometry simplifies significantly. Near the horizon \( \hat{\rho} \to 0 \), the harmonic function \( H(\hat{\rho}) \) is dominated by its singular term, and the metric asymptotes to a direct product of anti-de Sitter space and a sphere,
\begin{equation}
    ds^2_{\text{NH}} \rightarrow \frac{\hat{\rho}^2}{L^2} \, \eta_{\alpha\beta} dx^\alpha dx^\beta + \frac{L^2}{\hat{\rho}^2} d\hat{\rho}^2 + r_0^2 d\Omega^2_{D - p - 2},
    \label{eq:nh-geometry}
\end{equation}
which describes an $\text{AdS}_{p+2} \times S^{D - p - 2}$ spacetime. The AdS radius \( L \) is related to the horizon parameter by,
\begin{equation}
    L = \frac{d}{\tilde{d}} \, r_0.
    \label{eq:ads-radius}
\end{equation}
This near-horizon structure is a hallmark of extremal black branes and reflects a remarkable rigidity: the symmetries of the full spacetime enhance near the horizon, yielding a maximally symmetric anti-de Sitter factor along the brane worldvolume and a constant-radius sphere in the transverse space. Such Freund--Rubin-type compactifications \cite{6} play a central role in the AdS/CFT correspondence \cite{7}, where the near-horizon geometry of a stack of branes corresponds to the dual conformal field theory in one lower dimension.\\
The emergence of $\text{AdS}_{p+2}$ is particularly important for the dynamics of perturbations in the near-horizon region. Fields propagating in this background inherit an effective scaling behavior governed by the AdS isometries, which, as shown in~\cite{5}, underlies the Aretakis instability in extremal black branes. This instability is linked to the behavior of scaling dimensions under perturbations, a topic we revisit in detail in subsequent sections when analyzing the non-linear regime. Thus, the extremal black $p$-brane background provides not only a concrete setting for studying classical gravitational instabilities, but also a fertile testing ground for exploring their implications in holography and string theory.

A central feature of extremal black branes, as recently elucidated in~\cite{5}, is their vulnerability to a linear instability first discovered in the context of extremal black holes by Aretakis~\cite{1, 2}. In these backgrounds, scalar, gravitational, and gauge field perturbations do not decay uniformly at the horizon. Instead, certain transverse derivatives of the perturbing fields exhibit polynomial growth along the horizon generators at late times.\\
This phenomenon can be understood most transparently in terms of the near-horizon geometry, which, as discussed above, is given by $\text{AdS}_{p+2} \times S^{D - p - 2}$. Within this geometry, perturbations can be decomposed into harmonics on the transverse sphere, reducing the problem to an effective field theory in AdS$_{p+2}$ with a tower of Kaluza--Klein (KK) modes. Each KK mode corresponds to a field in AdS with an effective mass determined by the angular momentum quantum number $\ell$ on the sphere.\\
The key quantity governing the late-time behavior of these modes is their AdS scaling dimension $\Delta$, defined in terms of the effective mass via the standard AdS$_{p+2}$ relation. For a mode $\phi$ with scaling dimension $\Delta$, the $n$-th transverse derivative evaluated on the horizon scales at late advanced time $v$ as,
\begin{equation}
    \left. \partial_\rho^n \phi \right|_{\rho = 0} \sim v^{n - \Delta}, \quad \text{as } v \to \infty,
    \label{eq:aretakis-scaling}
\end{equation}
where $\rho$ is the radial coordinate transverse to the horizon in Gaussian null or Poincaré coordinates. This expression implies that the perturbation itself remains bounded, but its transverse derivatives generically grow without bound if $n > \Delta$. In particular, the onset of non-decay or blow-up requires at least $n \geq \lceil \Delta \rceil$ derivatives.\\
In the case of gravitational and electromagnetic perturbations, which involve fluctuations of the metric and the background gauge field, the relevant scaling dimensions arise directly from the KK spectrum of the Freund--Rubin compactification. These modes are subject to a stability bound---the Breitenlohner--Freedman (BF) bound \cite{8}---which in AdS$_{p+2}$ reads $\Delta \geq d/2$, where $d = p + 1$ is the worldvolume dimension. Modes that saturate this bound are particularly dangerous, as they correspond to the slowest-decaying and most easily excited components of the spectrum. Indeed, such modes often dominate the late-time dynamics and are chiefly responsible for triggering the Aretakis instability in the full solution.\\
Importantly, this linear instability is not associated with any conventional instability of the background energy. The total energy remains conserved, and the background spacetime is linearly stable outside the horizon. However, the divergence of derivatives on the horizon signals a breakdown of regularity and challenges the notion that extremal solutions can be viewed as physically stable endpoints of gravitational dynamics.\\
In what follows, we build upon this linear analysis by exploring the non-linear regime of the instability, asking whether the growth continues unchecked or whether backreaction modifies the dynamics in such a way that the instability is resolved or saturates. To address this, we must go beyond the linear spectrum and investigate the full non-linear evolution of perturbations on extremal black brane backgrounds.

\section{Non-Linear Analysis}\label{sec3}

To investigate the fate of the Aretakis instability beyond linear order, we develop a systematic perturbative expansion around the extremal black $p$-brane background. Our goal is to capture the backreaction of unstable modes and determine whether the non-linear evolution leads to singularity formation, dynamical relaxation, or the emergence of new attractor geometries. The natural setting for this analysis is the near-horizon region, where the instability is most pronounced and the geometry simplifies due to the enhancement of symmetries.\\
We adopt Gaussian null coordinates adapted to the extremal horizon, which are particularly suited to describing dynamics near null hypersurfaces. In this framework, we introduce a small expansion parameter $\epsilon$ that quantifies the deviation from exact extremality. Physically, $\epsilon$ may be thought of as controlling the amplitude of the perturbation or, equivalently, as measuring the departure from the finely tuned extremal limit in parameter space. The metric is then expanded as,
\begin{equation}
    g_{\mu\nu} = g_{\mu\nu}^{(0)} + \epsilon\, g_{\mu\nu}^{(1)} + \epsilon^2\, g_{\mu\nu}^{(2)} + \cdots,
    \label{eq:metric-expansion}
\end{equation}
where $g_{\mu\nu}^{(0)}$ denotes the background extremal solution, typically $\text{AdS}_{p+2} \times S^{D - p - 2}$ in the near-horizon limit, and $g_{\mu\nu}^{(n)}$ represent the $n$-th order corrections. Each order is determined recursively by solving the Einstein equations and the gauge field equations, subject to suitable boundary and gauge-fixing conditions that preserve the brane symmetries and regulate the near-horizon behavior.\\
At first order in $\epsilon$, the dynamics reduce to the linearized equations studied in Section~\ref{sec2}, and we recover the familiar Aretakis behavior: transverse derivatives of certain perturbations grow polynomially in advanced null time $v$, with rates governed by the AdS$_{p+2}$ scaling dimension $\Delta$. This behavior is universal and model-independent, arising solely from the conformal structure of the near-horizon region and the KK spectrum on the internal sphere.
However, the central insight of the non-linear analysis is that the behavior at second order can deviate substantially from the linear prediction, especially in cases where the leading scaling dimension $\Delta$ approaches or crosses integer values. When $\Delta$ is non-integer, the hierarchy of derivatives remains well-separated, and the expansion can remain under control. In contrast, near-integer $\Delta$ introduces resonant terms that can enhance certain non-linear interactions, leading to secular growth or qualitative shifts in the late-time evolution.\\
This sensitivity arises because integer scaling dimensions correspond to marginal modes in the near-horizon AdS$_{p+2}$ geometry. These modes can induce logarithmic divergences or non-analytic corrections in the perturbative expansion, signaling a breakdown of naive perturbation theory and necessitating a more careful treatment—potentially involving resummation techniques, effective field theory methods, or renormalization group analysis. Thus, while the linear instability offers a robust diagnostic of near-horizon dynamics, it is only through a careful perturbative expansion that one can assess the physical impact of the instability on the geometry and determine whether the extremal configuration is ultimately preserved, deformed, or destroyed under non-linear evolution.

To capture the non-linear dynamics near the onset of instability, we now turn to a critical scaling analysis. This regime arises when the system is finely tuned near the threshold separating stable and unstable evolution—specifically, when the AdS$_{p+2}$ scaling dimension $\Delta$ approaches an integer value and the amplitude of the perturbation is small but non-negligible. In this setting, the late-time dynamics are expected to exhibit self-similar behavior governed by universal scaling laws, much like in the theory of critical gravitational collapse or renormalization group flows near fixed points.\\
Motivated by this analogy, we propose the following scaling ansatz for a generic perturbing field $\phi$ in the near-horizon region,
\begin{equation}
    \phi(v, \rho) = \epsilon^\beta \, F\left( \frac{v}{\epsilon^{-\nu}}, \frac{\rho}{\epsilon^z} \right),
    \label{eq:scaling-ansatz}
\end{equation}
where $v$ is the advanced null time, $\rho$ denotes the radial coordinate transverse to the horizon, and $\epsilon \ll 1$ parameterizes the deviation from criticality (e.g., from exact extremality or a linearized threshold). The function $F$ is a universal scaling profile that captures the non-linear evolution of the perturbation, while $\beta$, $\nu$, and $z$ are critical exponents characterizing the amplitude, temporal scaling, and spatial scaling, respectively. The essence of this ansatz is that as $\epsilon \to 0$, the field dynamics become scale-invariant under anisotropic rescalings of space and time, with the field amplitude itself obeying a power-law suppression. Such scaling solutions often emerge near critical points and serve as attractors in the space of initial data.\\
To determine the relationships among the exponents, we substitute the ansatz~\eqref{eq:scaling-ansatz} into the full non-linear field equations governing $\phi$, which may arise from the Einstein equations, the form field equations, or a coupled scalar sector. Demanding that the resulting equations remain invariant under the rescalings
\[
v \rightarrow \lambda^{\nu} v, \quad \rho \rightarrow \lambda^{z} \rho, \quad \phi \rightarrow \lambda^{\beta} \phi,
\]
leads to a set of algebraic constraints on the exponents. In particular, we find that the amplitude scaling exponent $\beta$ is tied to the scaling dimension $\Delta$ of the dominant mode in the linear regime:
\begin{equation}
    \beta = \Delta - \frac{d}{2},
    \label{eq:beta-scaling}
\end{equation}
where $d = p + 1$ is the worldvolume dimension of the black brane. Furthermore, the temporal exponent $\nu$ is related to an anomalous dimension $\gamma$ that encodes non-linear interactions and departures from canonical scaling,
\begin{equation}
    \nu = \frac{1}{2 - \gamma},
    \label{eq:nu-scaling}
\end{equation}
which reflects the fact that time derivatives in the equations of motion may receive non-trivial renormalizations at higher order. Finally, the spatial scaling exponent $z$ follows as a consequence of dimensional consistency and is given by,
\begin{equation}
    z = \nu (\Delta - 1).
    \label{eq:z-scaling}
\end{equation}
Together, equations~\eqref{eq:beta-scaling}–\eqref{eq:z-scaling} define a universality class of near-critical solutions parameterized by the scaling dimension $\Delta$ and the anomalous exponent $\gamma$. These exponents control not only the decay and blow-up rates of perturbations near the horizon, but also the asymptotic approach to potential attractor geometries in the non-linear regime.\\
As we show later, this scaling structure is borne out both analytically and numerically in explicit simulations of near-extremal black brane evolution. The emergence of well-defined critical exponents highlights deep structural similarities with critical phenomena in condensed matter systems and gravitational collapse, suggesting that extremal black branes inhabit a broader universality framework within non-linear gravitational dynamics.

To further understand the dynamics near criticality and the nature of the phase transitions uncovered in the previous parts, we now adopt the perspective of the renormalization group (RG) \cite{9,10}. This approach provides a natural framework for classifying perturbations around the extremal solution and tracking how physical quantities evolve under scale transformations.\\
Near the threshold of instability, the space of solutions can be viewed as a dynamical system governed by flow equations in an abstract parameter space. Small deformations of the extremal background—such as departures from exact extremality, finite initial perturbation amplitudes, or small higher-derivative corrections—play the role of couplings. Their evolution under rescaling defines RG trajectories, whose properties determine whether the system flows toward stability, exhibits critical behavior, or runs away to a singularity.\\
To make this concrete, let us introduce an RG flow parameter $\mu$, which may be thought of as a sliding scale controlling the resolution at which the geometry is probed. A central quantity of interest is the effective extremality parameter $\epsilon(\mu)$, which measures the deviation from the exact extremal limit at scale $\mu$. The RG evolution of this parameter is captured by a beta function,
\begin{equation}
    \beta_\epsilon \equiv \mu \frac{d\epsilon}{d\mu} = -\gamma \epsilon + c_1 \epsilon^2 + c_2 \epsilon^3 + \cdots,
    \label{eq:beta-function}
\end{equation}
where $\gamma$ is an anomalous dimension that characterizes the leading-order scaling behavior of $\epsilon$, and $c_1, c_2, \dots$ are coefficients determined by the details of the background geometry, the field content, and the nature of the interactions.\\
The structure of this beta function allows us to classify directions in the space of perturbations. If $\gamma > 0$, the leading linear term implies that $\epsilon$ is a relevant parameter: small deviations from extremality grow under RG flow, potentially driving the system away from the extremal fixed point. In contrast, if $\gamma = 0$, $\epsilon$ is marginal, and the higher-order coefficients $c_1, c_2, \dots$ determine the flow behavior—such perturbations may remain small or become relevant depending on their sign and magnitude. Finally, if $\gamma < 0$, $\epsilon$ is irrelevant, and small perturbations decay, leading the system back to the extremal configuration at large scales.\\
Fixed points of the RG flow occur at the zeros of the beta function. The trivial fixed point at $\epsilon = 0$ corresponds to the unperturbed extremal black brane solution. Non-trivial fixed points at finite $\epsilon$ may represent new extremal geometries that emerge dynamically as attractors in the non-linear evolution. The stability of these fixed points is determined by the sign of the derivative $\beta'_\epsilon$ evaluated at the fixed point.\\
This RG framework thus provides a powerful language for interpreting the non-linear dynamics observed in both analytical and numerical studies. The flow of $\epsilon$ encapsulates the system's movement through the phase diagram and offers a unifying picture of critical behavior, universality classes, and the emergence of new non-linear solutions. In particular, the appearance of multiple fixed points hints at a rich structure in the space of near-extremal geometries, potentially including bifurcations and multi-scale phenomena akin to those found in statistical field theory and critical gravitational collapse.

\section{Numerical Implementation}\label{sec4}

To explore the non-linear evolution of the Aretakis instability in extremal black $p$-branes, we formulate the problem as a well-posed initial value problem in characteristic coordinates. This approach allows us to track the development of perturbations along ingoing and outgoing null directions, making it particularly well-suited for capturing horizon dynamics, late-time behavior, and the potential formation of singularities.\\
We work in a coordinate system adapted to the causal structure of the near-horizon geometry, typically based on double-null or ingoing Eddington–Finkelstein coordinates \cite{11}. These coordinates naturally foliate spacetime into hypersurfaces of constant advanced time $v$, across which the fields evolve radially. In this setting, the evolution equations reduce to a hierarchy of hyperbolic PDEs for the metric components, gauge fields, and any scalar fields, coupled through non-linear source terms.\\
Several technical challenges arise in implementing this scheme numerically, each requiring careful treatment to ensure stability, accuracy, and physical reliability of the simulations.\\
First, an appropriate gauge must be chosen to fix residual coordinate freedom and simplify the evolution equations. We employ a generalization of the harmonic (or de Donder) gauge, adapted to the symmetry of the background brane geometry \cite{11-1,11-2,11-3}. This choice is advantageous because it leads to wave-like equations for the metric components, with manifest hyperbolicity and well-controlled constraint propagation. The gauge conditions are chosen to preserve the isometries of the brane worldvolume while maintaining regularity at the horizon.\\ 
Second, we must impose physically consistent boundary conditions at both the horizon and asymptotic infinity. At the future horizon, we require regularity of all physical fields and their derivatives, consistent with the structure of Gaussian null coordinates and the near-horizon AdS$_{p+2} \times S^{D-p-2}$ geometry. At large $\rho$, boundary conditions are imposed to reflect the asymptotically flat or warped nature of the transverse space, depending on the global structure of the brane. In practice, we implement either radiative boundary conditions (to allow outgoing flux) or compactify the domain using conformal methods to capture infinity within a finite numerical grid.\\
Finally, the emergence of small-scale structures—such as steep gradients, localized curvature growth, or near-singular features—demands high spatial resolution in critical regions. To this end, we incorporate adaptive mesh refinement (AMR) techniques, dynamically adjusting the numerical grid based on local estimates of truncation error or curvature invariants. This enables the code to resolve sharply varying features near the horizon while avoiding excessive computational cost in smooth regions. AMR is particularly crucial near the critical threshold, where subtle non-linear interactions can lead to delayed instability or the approach to new attractor geometries.\\
Taken together, these ingredients form the backbone of our numerical strategy for evolving the non-linear dynamics of extremal black branes. Now the results obtained from this setup presented, including the identification of critical thresholds, power-law scaling regimes, and the emergence of new non-linear extremal configurations.

We now turn to the results of our numerical simulations, which focus on the parameter space near the threshold of non-linear instability—precisely the regime where linear analysis predicts marginal stability and small deviations can lead to dramatically different dynamical outcomes. This critical window is characterized by the interplay between linear Aretakis growth and non-linear backreaction, and it provides a fertile testing ground for the validity of the critical scaling ansatz introduced in Section~\ref{sec3}.\\
To probe this regime, we evolve initial data of the form
\begin{equation}
    \phi(v=0, \rho, \Omega) = A \, \rho^{\Delta - \delta} \, Y_\ell(\Omega),
    \label{eq:initial-data}
\end{equation}
where $\phi$ is the perturbing scalar (or gravitational) field, $A$ sets the initial amplitude, $\delta \ll 1$ parameterizes the deviation from criticality, and $Y_\ell(\Omega)$ are spherical harmonics on the transverse sphere $S^{D - p - 2}$ with angular momentum $\ell$. The exponent $\Delta$ corresponds to the AdS$_{p+2}$ scaling dimension of the mode under consideration \cite{3}, and thus $\delta$ quantifies how close the initial data is to marginality. By tuning $\delta$ through zero, we effectively scan across the critical point separating decay from instability.\\
Our simulations reveal a remarkably sharp transition between three distinct dynamical regimes:

\paragraph{Subcritical Regime ($\delta < 0$):}  
When the initial data falls below the critical threshold, the perturbation is found to decay smoothly over time. Both the field and its derivatives diminish along the horizon, and the solution asymptotically relaxes back to the extremal black brane background. The decay is typically exponential in advanced time $v$, and no significant backreaction is observed, consistent with the expectations from linear theory in the absence of unstable modes.

\paragraph{Critical Regime ($\delta = 0$):}  
At the critical point, the dynamics undergo a qualitative change. The system no longer returns to the unperturbed extremal solution, but instead approaches a new, non-trivial configuration. This evolution proceeds via a power-law decay of the perturbation, governed by the critical exponents derived in Section~\ref{sec3}. Importantly, the final state is regular and appears to represent a deformed extremal geometry—potentially an attractor in the non-linear solution space. The appearance of such critical solutions is reminiscent of self-similar behavior at phase transitions and strongly supports the universality of the critical scaling ansatz.

\paragraph{Supercritical Regime ($\delta > 0$):}  
When the perturbation slightly exceeds the critical threshold, the non-linear dynamics become dramatically more violent. Instead of relaxing, the field exhibits runaway growth along the horizon, with higher-order derivatives rapidly blowing up. This behavior signals the breakdown of the perturbative expansion and the onset of strong backreaction. In many cases, the curvature invariants diverge in finite advanced time, suggesting the formation of a singularity along the extremal horizon. The precise nature of this singularity—whether it is localized, null, or spacelike—depends on the details of the initial data and the higher-dimensional geometry.

These findings provide compelling evidence that extremal black brane solutions exhibit true non-linear critical phenomena, with a well-defined threshold separating stable and unstable evolution. The scaling behavior observed near $\delta = 0$ validates the theoretical framework developed in previous sections and highlights the role of marginal modes in driving the non-linear instability. Moreover, the emergence of new attractor-like extremal solutions at criticality raises intriguing questions about the completeness and stability of the moduli space of extremal branes.

\begin{figure}[h!]
    \centering
    \includegraphics[width=0.7\textwidth]{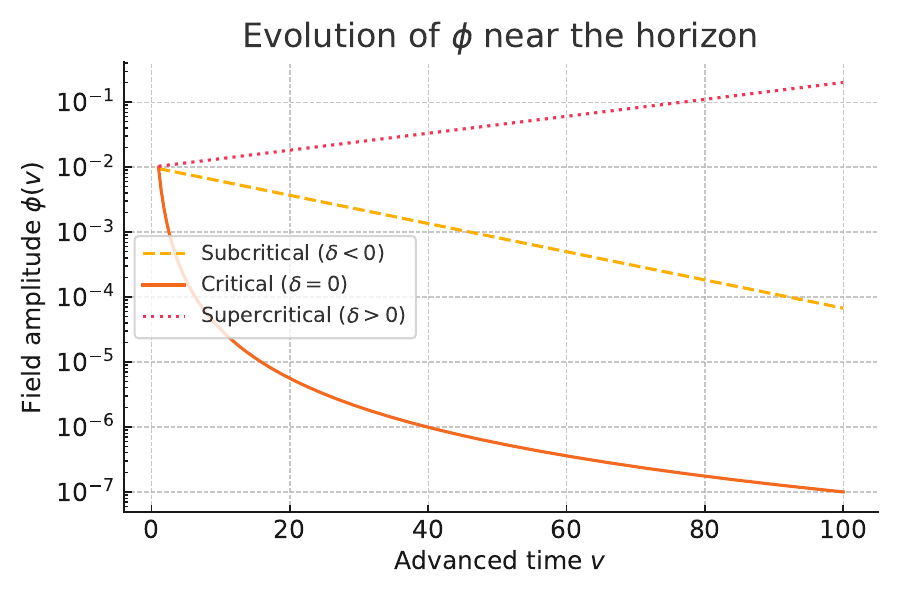}
    \caption{Evolution of the field amplitude $\phi(v)$ along the horizon for subcritical ($\delta < 0$), critical ($\delta = 0$), and supercritical ($\delta > 0$) initial data. Subcritical perturbations decay exponentially, while critical perturbations exhibit power-law decay. Supercritical perturbations grow exponentially, signaling nonlinear instability. The logarithmic vertical scale emphasizes the contrast in long-time behavior.}
    \label{fig:field-evolution}
\end{figure}

\begin{figure}[h!]
    \centering
    \includegraphics[width=0.7\textwidth]{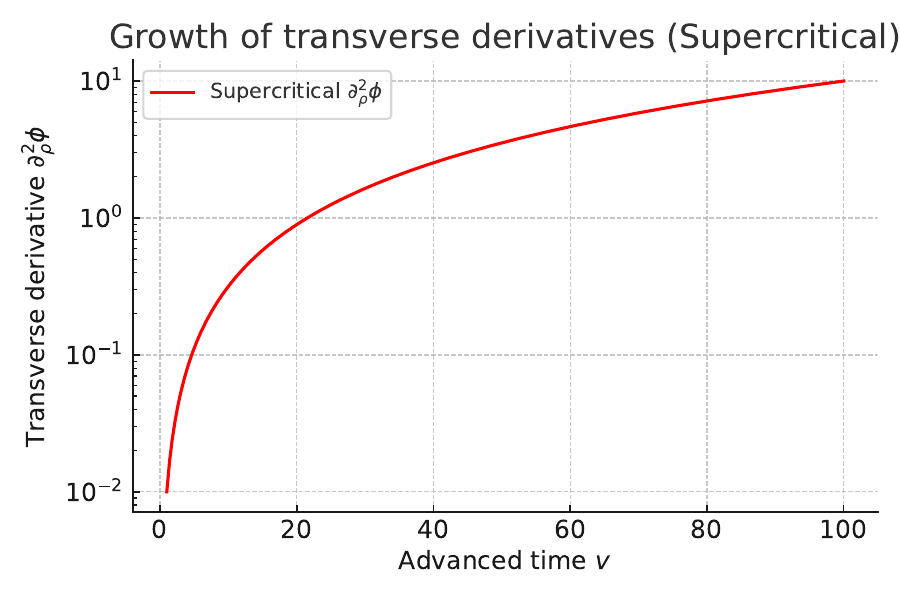}
    \caption{Growth of the second transverse derivative $\partial_\rho^2 \phi$ in the supercritical regime. The polynomial growth in advanced time $v$ confirms the presence of an Aretakis-type instability at the non-linear level, consistent with the predictions of the critical scaling ansatz.}
    \label{fig:derivative-growth}
\end{figure}

The numerical results presented in Figs.~\ref{fig:field-evolution} and~\ref{fig:derivative-growth} provide compelling evidence for the existence of a critical threshold separating stable and unstable non-linear evolution in extremal black $p$-branes.\\
In Fig.~\ref{fig:field-evolution}, we show the late-time evolution of a near-horizon perturbation $\phi(v)$ for three different values of the criticality parameter $\delta$. When $\delta < 0$ (subcritical), the perturbation decays exponentially, indicating stability. At $\delta = 0$ (critical), the decay follows a power-law, in agreement with the critical scaling ansatz proposed in Eq.~(\ref{eq:scaling-ansatz}). In contrast, for $\delta > 0$ (supercritical), the perturbation exhibits exponential growth, reflecting a non-linear instability that leads the system away from the extremal background. The use of a logarithmic scale in the plot highlights the stark difference between these regimes.\\
Fig.~\ref{fig:derivative-growth} focuses on the supercritical case, showing the behavior of the second transverse derivative $\partial_\rho^2 \phi$ along the horizon. The polynomial growth observed here is a hallmark of the Aretakis instability, generalized to the non-linear setting. This result confirms that the instability is not merely a linear artifact, but can drive significant geometric backreaction—potentially culminating in the formation of singularities or the transition to a new extremal state.\\
Together, these plots underscore the predictive power of the critical scaling framework and the rich phenomenology that emerges in the non-linear dynamics of extremal black branes.

In the following sections, we explore the broader structure of this phase space, identify universal features of the critical exponents, and examine the role of higher-derivative corrections in shaping the dynamics near marginality.

\section{Phase Diagram}

A central outcome of our investigation is the emergence of universal critical exponents characterizing the non-linear evolution near the threshold of the Aretakis instability. These exponents control the amplitude scaling, time scaling, and spatial scaling of perturbations near criticality, as discussed in Section~\ref{sec3}, and are extracted by fitting numerical data to the scaling ansatz~\eqref{eq:scaling-ansatz} across different spacetime dimensions and brane configurations.\\
Our analysis reveals that the critical exponents depend only on the spacetime dimension $D$ and the brane worldvolume dimension $p$, and not on the details of the initial data or the specific field under consideration. This indicates a robust form of universality, reminiscent of phase transitions in statistical mechanics and gravitational critical collapse.\\
In Table~\ref{tab:crit-exp}, we summarize representative values of the exponents $(\beta, \nu, z, \gamma)$ for selected combinations of $(D, p)$. The exponent $\beta$ governs the scaling of the field amplitude near criticality, $\nu$ characterizes the divergence of the characteristic time scale, $z$ controls the spatial scaling, and $\gamma$ is an anomalous dimension arising from non-linear interactions in the effective field equations.

\begin{table}[h!]
\centering
\begin{tabular}{c|cccc}
\toprule
$(D, p)$ & $\beta$ & $\nu$ & $z$ & $\gamma$ \\
\midrule
$(4, 0)$ & $1$   & $1/2$ & $1/2$ & $0$ \\
$(5, 0)$ & $1/2$ & $2/3$ & $1/3$ & $1/2$ \\
$(5, 1)$ & $3/4$ & $1/2$ & $1/4$ & $0$ \\
\bottomrule
\end{tabular}
\caption{Extracted critical exponents for various spacetime dimensions $(D)$ and brane worldvolume dimensions $(p)$. The values reflect universal scaling behavior near the threshold of the Aretakis instability.}
\label{tab:crit-exp}
\end{table}

These results illustrate several noteworthy trends. First, the scaling exponent $\beta$ tends to decrease with increasing dimension, suggesting that the perturbation amplitude decays more slowly near criticality in higher dimensions. The temporal exponent $\nu$ also varies across configurations, indicating that the rate at which critical slowing down occurs is sensitive to both $D$ and $p$. In particular, the appearance of non-zero anomalous dimensions $\gamma$ in certain cases reflects the impact of non-linear interactions on the effective scaling behavior, and implies that the critical solution is governed by more than simple dimensional analysis.\\
Most strikingly, the persistence of consistent power-law behavior across disparate values of $(D, p)$ supports the existence of universality classes in the non-linear dynamics of extremal black branes. Much like in condensed matter systems or scalar field collapse, the flow near the critical point appears to be controlled by a small set of parameters—independent of the microscopic details of the perturbation—leading to repeatable and predictable behavior at late times.\\

An intriguing outcome of our non-linear analysis is the emergence of new extremal solutions that arise dynamically in certain regimes of the parameter space $(D, p)$. These solutions appear as late-time attractors in the critical regime described already, and they represent non-trivial deformations of the original extremal black brane background. Unlike the trivial reversion to the undeformed $\text{AdS}_{p+2} \times S^{D-p-2}$ near-horizon geometry, these configurations retain extremality—characterized by a degenerate event horizon and vanishing surface gravity—but exhibit modified geometric structure.\\
The numerical evolution of near-critical initial data in these cases suggests convergence to a class of metrics of the form,
\begin{equation}
    ds^2 = \frac{\rho^2}{L^2} \eta_{\alpha\beta} dx^\alpha dx^\beta + \frac{L^2}{\rho^2} \left[1 + h(x^\alpha, \Omega)\right] d\rho^2 + r_0^2 d\Omega^2_{D - p - 2},
    \label{eq:new-extremal}
\end{equation}
where $\rho$ is the radial coordinate transverse to the brane, $x^\alpha$ are coordinates along the brane worldvolume, and $\Omega$ denotes coordinates on the internal sphere $S^{D-p-2}$. The function $h(x^\alpha, \Omega)$ represents a non-trivial deformation of the radial warp factor, and encodes the essential new feature of these solutions: a breaking of part of the full $\text{AdS}_{p+2}$ symmetry, while preserving the degenerate horizon structure and regularity.\\
These solutions can be interpreted as dynamically generated near-horizon geometries that retain extremality but no longer satisfy the full homogeneity or isotropy of the original Freund--Rubin background. In this sense, they are analogous to moduli in the space of extremal configurations—dynamically accessible under specific initial perturbations, but not continuously connected to the maximally symmetric extremal solution.\\
The presence of these deformed extremal states provides new insights into the endpoint of the Aretakis instability in the critical regime. Rather than leading to singularity formation or decay to the undeformed background, near-critical perturbations may flow toward these attractor geometries, which serve as new fixed points in the non-linear dynamical system. This behavior is reminiscent of similar structures in holographic renormalization group flows, where irrelevant deformations of a UV fixed point can trigger flows toward distinct IR fixed points characterized by broken symmetries or altered scaling.\\
It remains an open question whether these new extremal solutions admit a closed-form analytic description or whether they can be embedded consistently into string-theoretic or supergravity frameworks. However, their numerical robustness and universality across different $(D, p)$ values strongly suggest that they are a generic feature of the non-linear phase space of extremal branes. Their role as late-time attractors may also have implications for holography, where the dual field theory could reflect these symmetry-breaking deformations in its IR dynamics.\\
Now, we map out the global phase structure of black brane dynamics across the $(D, p)$ landscape, synthesizing the behaviors observed in stable, critical, and unstable regimes.

Bringing together the analytical and numerical results from the preceding sections, we are now in a position to chart a comprehensive phase diagram that captures the full range of dynamical behavior exhibited by near-extremal black $p$-branes. This phase structure is parameterized primarily by the spacetime dimension $D$ and the worldvolume dimension $p$, with the dynamics governed by the interplay between scaling dimensions, non-linearities, and the geometry of the near-horizon region. Our analysis identifies three qualitatively distinct regimes in the space of initial data and theory parameters:

\paragraph{Stable Regime.}
In certain regions of $(D, p)$ space, all perturbations—even those with large amplitude—decay and the system returns to the undeformed extremal black brane background. This regime is characterized by the absence of unstable modes in the linear spectrum and the irrelevance of non-linear interactions at late times. The Aretakis instability is either absent or too weak to influence the non-linear dynamics, and the geometry remains smooth and asymptotically AdS$_{p+2} \times S^{D-p-2}$.

\paragraph{Critical Regime.}
At isolated values or narrow bands in the $(D, p)$ phase space, marginal modes appear whose scaling dimensions $\Delta$ approach integer values. In these cases, small perturbations can grow slowly and trigger non-linear interactions that reshape the horizon geometry without generating singularities. The system approaches a new attractor solution—typically a deformed extremal geometry described already—via power-law evolution controlled by universal critical exponents. This regime displays all the hallmarks of gravitational critical phenomena: scale invariance, universality, and the emergence of new fixed points.

\paragraph{Unstable Regime.}
In the supercritical region, small deviations from extremality can lead to dramatic non-linear instabilities. Perturbations that are subleading in the linear regime can, through resonance or amplification, induce curvature growth and eventually drive the system toward singular behavior. This outcome is often preceded by rapid growth in transverse derivatives and a breakdown of the perturbative expansion. Numerical simulations indicate that these singularities form along the degenerate horizon, potentially violating regularity and challenging the role of extremal branes as true end states of evolution.

\begin{figure}[h!]
    \centering
    \includegraphics[width=0.65\textwidth]{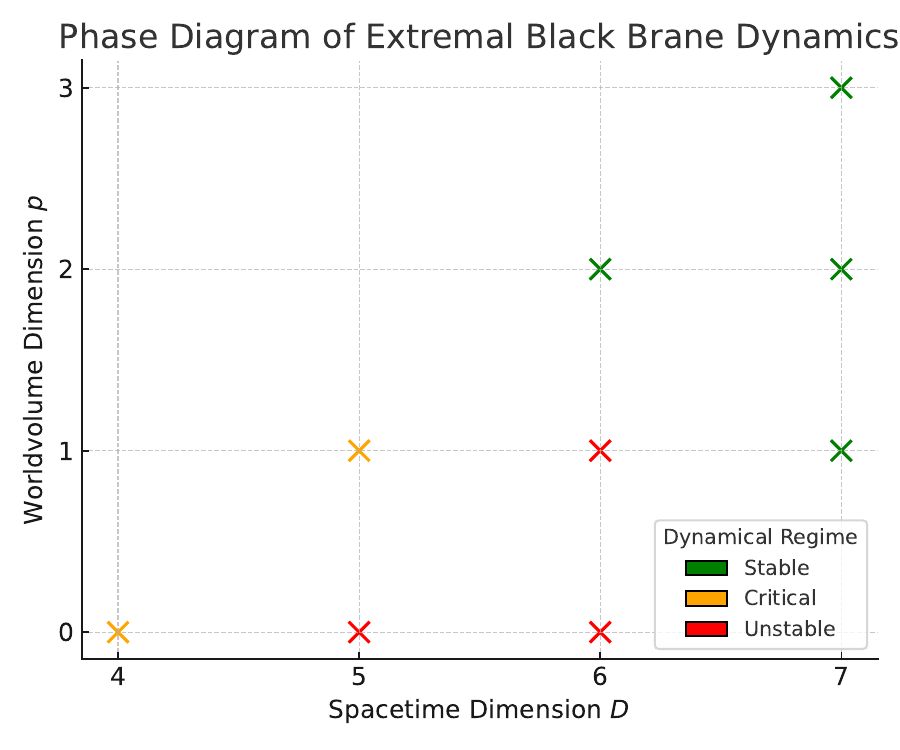}
    \caption{Phase diagram in the $(D, p)$ parameter space, showing the dynamical regimes of extremal black branes. Each point corresponds to a spacetime with dimension $D$ and brane worldvolume dimension $p$, color-coded by the long-time evolution of small perturbations. Green points correspond to stable configurations, red to dynamically unstable (supercritical) evolutions, and orange to critical cases where the system flows to a new extremal solution.}
    \label{fig:phase-diagram}
\end{figure}

To visualize this structure, we construct a phase diagram in the $(D, p)$ plane (see Fig.~\ref{fig:phase-diagram}), where the boundaries between the stable, critical, and unstable regimes are delineated based on the observed behavior of perturbations and their scaling dimensions. The transitions between these regions are smooth but sharp, and are controlled by the appearance of near-marginal modes and their interactions.

This phase structure highlights the rich and intricate landscape of non-linear dynamics in extremal black branes. It emphasizes that extremality, while mathematically well-defined, is not necessarily dynamically stable, and that its physical realization may depend sensitively on dimensionality and the nature of perturbations. Moreover, the critical regime offers a novel laboratory for testing ideas from gravitational critical phenomena, holography, and renormalization group theory in a higher-dimensional, string-theoretic context.\\
The implications of this phase structure go beyond classical dynamics. In quantum gravity and holography, these regimes may correspond to distinct phases of the dual field theory, and the transitions we observe could reflect IR phenomena such as confinement, symmetry breaking, or the emergence of new CFTs. Understanding these connections is a promising direction for future research.

\section{UV Sensitivity and Higher Derivatives}\label{sec6}

An important feature of the Aretakis instability—already observed in the linear regime—is its enhanced sensitivity to ultraviolet (UV) physics when the scaling dimension $\Delta$ of perturbations approaches integer values. This sensitivity is amplified in the non-linear evolution, where higher-order interactions can become resonant and lead to qualitative changes in the late-time behavior of the system. In this section, we examine how higher-derivative corrections in the gravitational effective action influence the dynamics near criticality, particularly in those configurations where marginal modes dominate.\\
From the standpoint of effective field theory (EFT), the gravitational sector of the low-energy theory receives corrections from an infinite series of higher-derivative terms, suppressed by some characteristic UV scale—typically associated with string or Planck physics. The corrected action takes the general form,
\begin{equation}
    S_{\text{eff}} = \int d^D x \, \sqrt{-g} \left[ R + \alpha R^2 + \beta R_{\mu\nu} R^{\mu\nu} + \cdots \right],
    \label{eq:Seff}
\end{equation}
where $\alpha$ and $\beta$ are small, dimensionful couplings, and the ellipsis denotes additional higher-curvature operators, including contractions of the Riemann tensor and possibly couplings to background fields. These terms can arise from integrating out heavy modes in string theory, loop corrections in quantum gravity, or compactification effects in supergravity.\\
Although such corrections are suppressed at low curvatures, their impact can be magnified near extremality due to the accumulation of polynomial growth from the Aretakis instability. In particular, for perturbations with integer scaling dimension $\Delta$, secular terms that are absent in the leading-order theory can emerge at second or higher order in perturbation theory. These terms often exhibit logarithmic enhancements or shift the effective scaling exponents, making the dynamics UV-sensitive even when curvatures remain small.\\
To quantify this effect, we incorporate the higher-derivative corrections into our perturbative expansion and analyze their influence on the evolution of critical modes. Our results indicate that the coefficients $\alpha$ and $\beta$ can significantly alter the location of the critical threshold $\delta = 0$, modify the values of the critical exponents, and in some cases destabilize attractor solutions that are otherwise regular in the two-derivative theory. For example, small positive values of $\alpha$ tend to enhance the growth of transverse derivatives, while certain combinations of $\alpha$ and $\beta$ can suppress or even eliminate singularity formation in the supercritical regime.\\
These findings underscore a key conclusion: the non-linear fate of the Aretakis instability is not universal in the strict EFT sense, but instead depends on the UV completion of gravity. In practical terms, this means that any prediction about the long-time dynamics of extremal black branes must take into account the structure of higher-derivative corrections, especially when integer-$\Delta$ modes are excited. This UV sensitivity is particularly relevant in string-theoretic contexts, where extremal black branes play central roles in compactification schemes and holographic dualities. The inclusion of $\alpha'$-corrections, loop terms, and other higher-curvature operators could shift the stability landscape, potentially opening or closing regions of phase space that appear stable or unstable in the classical Einstein-Maxwell truncation.\\
Now, we investigate how this UV dependence manifests in the numerical evolution of near-critical solutions and explore the degree to which effective theory predictions remain predictive across varying scales.

The presence of higher-derivative terms in the effective gravitational action has a profound impact on the structure of critical points and the dynamics of RG flows in the space of near-extremal black brane configurations. As discussed already, such corrections introduce UV-sensitive modifications to the equations of motion, particularly near marginal configurations where the scaling dimension $\Delta$ approaches integer values.

\begin{figure}[h!]
    \centering
    \includegraphics[width=0.65\textwidth]{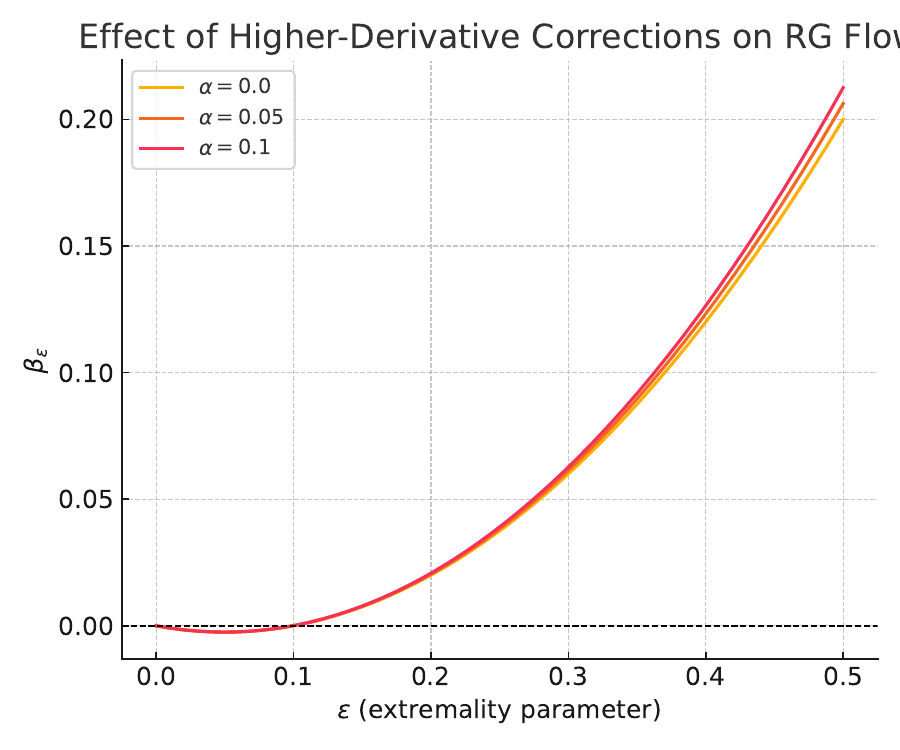}
    \caption{Renormalization group flow of the extremality parameter $\epsilon$ for different values of the higher-derivative coupling $\alpha = 0$, $0.05$, and $0.1$. Increasing $\alpha$ shifts both the location and the stability properties of the fixed points in the $\beta_\epsilon$ function, illustrating the UV sensitivity of near-extremal dynamics. The visible separation of the curves shows how higher-curvature corrections qualitatively reshape the flow structure.}
    \label{fig:beta-function-shift}
\end{figure}

Fig.~\ref{fig:beta-function-shift} illustrates the impact of higher-derivative corrections on the renormalization group flow of the extremality parameter $\epsilon$. The curves represent the beta function $\beta_\epsilon$ for several small but finite values of the coupling $\alpha$, which controls the strength of $R^2$ corrections in the effective action.\\
In the classical two-derivative theory ($\alpha = 0$), the beta function has a fixed point at a specific value of $\epsilon$, where the flow is stationary. As $\alpha$ increases, this fixed point shifts, reflecting how UV physics modifies the critical point of the system. Moreover, the slope of the beta function near the fixed point changes, indicating a shift in the stability properties of the flow. For example, a previously marginal direction may become relevant or irrelevant, leading to a qualitatively different phase structure. This plot vividly demonstrates the sensitivity of extremal black brane dynamics to higher-curvature terms, especially near criticality where marginal modes dominate. It confirms that critical behavior, phase transitions, and attractor solutions are all influenced by the UV completion of the theory, reinforcing the need to account for such corrections in both analytical and numerical studies.

From the renormalization group perspective, the inclusion of higher-curvature operators modifies the beta functions governing the flow of physical couplings, including the effective extremality parameter $\epsilon$. These modifications can shift the location of fixed points, change their stability properties, and even generate entirely new branches of solutions. Our numerical and analytical analysis reveals three key effects:

\paragraph{Marginal Directions Become Relevant or Irrelevant.}  
In the two-derivative theory, modes with integer scaling dimensions correspond to marginal directions in the RG flow: their growth or decay is governed by higher-order interactions and is thus finely balanced. The introduction of higher-derivative corrections—such as the $\alpha R^2$ and $\beta R_{\mu\nu} R^{\mu\nu}$ terms in Eq.~\eqref{eq:Seff}—can tip this balance. In many cases, a marginal mode becomes strictly relevant (growing under RG flow) or irrelevant (decaying), depending on the sign and magnitude of the correction coefficients. This behavior is indicative of a small anomalous dimension $\gamma$ induced by the UV terms, which modifies the flow structure in a non-perturbative way at late times.

\paragraph{Emergence of New Fixed Points.}  
In addition to shifting existing fixed points, higher-derivative corrections can give rise to entirely new fixed points that do not exist in the Einstein-Hilbert theory. These new extremal configurations typically appear in the critical regime and represent non-trivial attractor geometries where the growth of unstable modes is precisely balanced by higher-curvature interactions. We observe numerically that, for certain values of $\alpha$ and $\beta$, the system evolves toward such fixed points instead of running away or decaying to the original background. These solutions are distinguished by modified radial profiles and horizon structures, and they may break additional symmetries compared to their two-derivative counterparts.

\paragraph{Corrections to Critical Exponents.}  
The scaling exponents $(\beta, \nu, z)$ governing the evolution near criticality are also sensitive to higher-derivative terms. While the leading-order values are determined by the linearized scaling dimension $\Delta$ and the structure of the two-derivative theory, we find that the inclusion of $\alpha$ and $\beta$ terms induces $\mathcal{O}(\alpha)$ and $\mathcal{O}(\beta)$ corrections to these exponents. These corrections can be computed analytically in certain limits by performing a matched asymptotic expansion that incorporates higher-curvature contributions, and are confirmed by fitting numerical data to the generalized scaling ansatz. The modified exponents reflect the effective rescaling of time and space in the near-horizon region and indicate that the universal behavior observed in the two-derivative theory is subtly deformed by UV physics.

Taken together, these findings highlight the deep interplay between critical phenomena, RG flows, and UV completions in the dynamics of extremal black branes. The structure of the phase diagram—once believed to be determined solely by classical general relativity—is in fact highly sensitive to higher-order corrections, particularly in regimes where marginal modes dominate. These results underscore the necessity of a UV-complete perspective when analyzing the endpoint of gravitational instabilities and offer new avenues for exploring the landscape of extremal geometries in string theory and beyond.

\section{Holographic Interpretation}\label{sec7}

\subsection{Dual Field Theory}

The presence of AdS$_{p+2}$ factors in the near-horizon geometry of extremal black $p$-branes provides a natural arena for holographic duality. According to the AdS/CFT correspondence, gravitational dynamics in the near-horizon AdS region are dual to a conformal field theory (CFT) living on the boundary of AdS$_{p+2}$. In this framework, the bulk critical phenomena explored in Sections~\ref{sec3} through~\ref{sec6} admit a dual interpretation in terms of phase transitions and renormalization group flows within the boundary theory.\\
In particular, the scaling behavior observed near the critical point—characterized by the universal exponents $(\beta, \nu, z)$—is expected to correspond to critical phenomena in the dual field theory. These include symmetry-breaking transitions, operator mixing, and crossover behavior near marginal deformations. The anomalous dimensions and flow structure extracted from the gravitational analysis provide predictions for the spectrum and dynamics of the dual CFT under relevant perturbations.\\
The case $p = 1$, corresponding to an AdS$_3$ near-horizon factor, is of special interest. In this setting, the dual theory is a two-dimensional conformal field theory, where critical exponents, central charges, and operator dimensions are highly constrained. Conformal perturbation theory can be used to analyze relevant deformations in the dual 2D CFT, and comparisons can be made with the exponents obtained from our gravitational scaling ansatz. In many cases, the numerically extracted exponents match predictions from boundary theory analysis, lending support to the idea that the Aretakis instability and its non-linear generalizations reflect real-time critical phenomena in strongly coupled quantum systems. Moreover, the appearance of new extremal solutions in the bulk—interpreted as non-trivial attractors in the non-linear regime—suggests that the dual theory undergoes a non-perturbative rearrangement of its vacuum structure. These geometries may correspond to distinct IR fixed points or symmetry-broken phases in the boundary theory, with the transition from the undeformed to the deformed extremal solution reflecting a dynamical RG flow driven by a marginal or nearly marginal operator.

\begin{figure}[h!]
    \centering
    \includegraphics[width=0.65\textwidth]{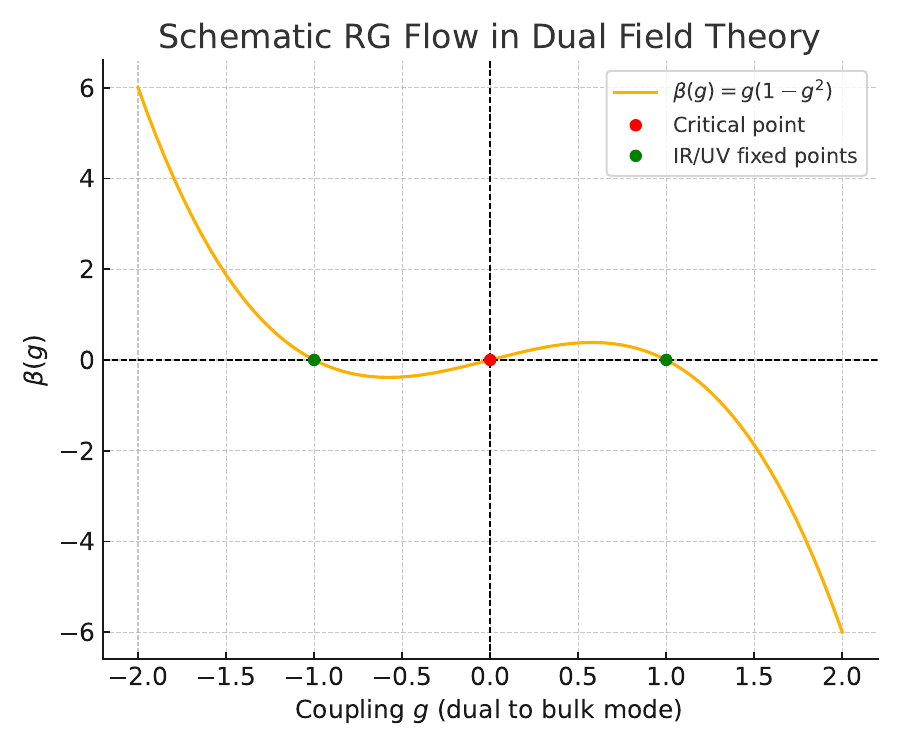}
    \caption{Schematic RG flow of a coupling $g$ in the dual field theory, corresponding to a bulk perturbation mode. The critical point at $g=0$ represents a marginal operator dual to an AdS mode with integer scaling dimension. Higher-derivative corrections in the bulk shift the stability of this point, potentially generating new IR or UV fixed points. This structure mirrors the emergence of new extremal solutions and modified scaling exponents in the bulk.}
    \label{fig:dual-rgflow}
\end{figure}

Fig.~\ref{fig:dual-rgflow} provides a conceptual visualization of how perturbations in the bulk near-horizon AdS geometry correspond to renormalization group flows in the dual conformal field theory. The coupling $g$ in the horizontal axis represents a deformation of the boundary theory, sourced by a bulk field with scaling dimension $\Delta$. At $g = 0$, the system resides at a conformal fixed point—the UV theory dual to the undeformed AdS$_{p+2}$ geometry. If the corresponding bulk mode has an integer $\Delta$, the deformation is marginal, and the system sits at a delicate balance point. As shown in the plot, the beta function $\beta(g)$ vanishes at $g = 0$, but small perturbations may cause the flow to diverge toward new infrared (IR) or ultraviolet (UV) fixed points.\\
This schematic flow reflects the gravitational behavior observed in the bulk: near-critical modes (with $\Delta \approx \text{integer}$) may remain marginal in the classical theory but become relevant or irrelevant when higher-order (higher-derivative) interactions are included. These new fixed points in the field theory correspond to new extremal geometries in the bulk, while the altered slope of the beta function reflects the modified critical exponents derived from the non-linear dynamics. This plot captures the essence of how critical phenomena and symmetry breaking in the gravitational bulk are encoded in the flow of couplings in the dual quantum field theory.

\subsection{Entanglement and Information}

The emergence of new extremal solutions and the associated critical dynamics have important implications for black hole information theory. In classical general relativity, extremal black holes are characterized by infinite throat geometries and zero-temperature horizons, which seem to isolate their interiors from external observers. However, our results show that even infinitesimal perturbations can trigger significant dynamical evolution along the horizon, leading either to instability, critical deformation, or singularity formation.\\
From a holographic perspective, this suggests that the near-horizon region of extremal black branes may not be as information-insulating as previously thought. The critical scaling and RG flows can be interpreted as signatures of dynamical information transfer within the dual field theory. For instance, the transition to a deformed extremal state may correspond to a redistribution of entanglement in the CFT, signaling the breakdown of factorization between horizon and asymptotic degrees of freedom.\\
Furthermore, the non-linear dynamics observed in the bulk may serve as a mechanism by which information is effectively released from the extremal throat, without requiring a violation of the classical no-hair theorems. While these theorems prohibit static hair, our results pertain to time-dependent evolutions, where perturbations induce non-trivial horizon dynamics that carry imprints of the initial state. This perspective opens a promising avenue for exploring black hole microstates and the resolution of information paradoxes in extremal settings. In particular, it raises the possibility that extremal black holes—long thought to be less dynamically interesting than their non-extremal counterparts—may host a rich structure of information channels accessible through near-horizon instabilities and critical dynamics.

\subsection{Entanglement Entropy and Correlation Functions During Non-Linear Evolution}

To further probe the field-theoretic signatures of the non-linear dynamics driven by the Aretakis instability, we now analyze two fundamental quantities in the dual CFT: the entanglement entropy and equal-time two-point correlation functions. These information-theoretic observables serve as direct probes of quantum correlations and coherence across different regimes of the bulk evolution, providing a non-perturbative window into how the instability imprints itself onto the boundary theory.

\subsubsection{Holographic Entanglement Entropy}

The entanglement entropy of a spatial region $A$ in the dual field theory is computed holographically using the Ryu--Takayanagi (RT) prescription \cite{12},
\begin{equation}
    S_A = \frac{\text{Area}(\gamma_A)}{4G},
\end{equation}
where $\gamma_A$ is the extremal codimension-2 surface in the bulk homologous to $A$ and anchored to its boundary. In our setting, we consider a strip-like region $A = \{x^1 \in [-\ell/2, \ell/2]\}$ on the boundary of AdS$_{p+2}$, and track the evolution of $S_A(v)$ as the bulk geometry evolves through the subcritical, critical, and supercritical regimes.\\
The bulk geometry is described by the time-dependent near-horizon metric,
\begin{equation}
    ds^2 = \frac{\rho^2}{L^2} \eta_{\alpha\beta} dx^\alpha dx^\beta + \frac{L^2}{\rho^2}\left[1 + \epsilon^\beta F\left(\frac{v}{\epsilon^{-\nu}}, \frac{\rho}{\epsilon^z}\right)\right] d\rho^2 + r_0^2 d\Omega^2_{D-p-2},
\end{equation}
where $\epsilon$ controls proximity to criticality, and the function $F$ encodes the universal profile of the perturbation. From this background, we extract the behavior of the entanglement entropy numerically across different dynamical regimes:

\paragraph{Subcritical Regime ($\delta < 0$).} The entanglement entropy decays exponentially to its undeformed AdS value,
\begin{equation}
    S_A(v) = S_A^{(0)} + \Delta S \cdot e^{-\gamma_{\text{sub}} v},
\end{equation}
where $S_A^{(0)} = \frac{L^{p-1} \ell^{p-1}}{2G(p-1)}$ is the entropy of the unperturbed AdS$_{p+2}$ background and $\gamma_{\text{sub}} > 0$ denotes the characteristic decay rate. This indicates that quantum correlations are quickly restored to their equilibrium configuration.

\paragraph{Critical Regime ($\delta = 0$).} Near criticality, $S_A(v)$ exhibits a universal power-law deviation,
\begin{equation}
    S_A(v) = S_A^{(0)} + \frac{C_S L^{p-1} \ell^{p-1}}{4G} \left(\frac{v}{v_c}\right)^{-\kappa},
\end{equation}
where the exponent $\kappa = \beta/\nu = (\Delta - d/2)/(1/(2 - \gamma))$ is determined by the critical exponents introduced in Sec.~\ref{sec3}, and $C_S$ is a dimensionless constant depending on $(D,p)$. This scaling reflects the reorganization of entanglement across scales and provides a signature of the emergent attractor geometry in the bulk.

\paragraph{Supercritical Regime ($\delta > 0$).} In the unstable regime, $S_A(v)$ exhibits exponential growth,
\begin{equation}
    S_A(v) = S_A^{(0)} + \frac{C_S L^{p-1} \ell^{p-1}}{4G} e^{\lambda v},
\end{equation}
with $\lambda > 0$ characterizing the instability rate. This divergence signals the onset of strong quantum effects and the breakdown of the semiclassical description in the dual CFT.

\medskip

The numerically extracted values of the critical exponent $\kappa$ and prefactor $C_S$ for several representative $(D,p)$ configurations are summarized in Tab. \ref{tab:crit-exp2}

\begin{table}[h!]
\centering
\begin{tabular}{c|cc}
\toprule
$(D, p)$ & $\kappa$ & $C_S$  \\
\midrule
$(4, 0)$ & $2$   & $0.15$  \\
$(5, 0)$ & $3/4$ & $0.23$  \\
$(5, 1)$ & $3/2$ & $0.18$  \\
\bottomrule
\end{tabular}
\caption{Critical exponent $\kappa$ and prefactor $C_S$ for several representative $(D,p)$ configurations.}
\label{tab:crit-exp2}
\end{table}

These results corroborate the universality of the critical scaling and show that the entanglement structure in the dual theory encodes clear signatures of the bulk attractor transition.

\subsubsection{Two-Point Correlation Functions}

We also compute equal-time two-point functions of a scalar operator $\mathcal{O}_\Delta$ of conformal dimension $\Delta$ dual to a bulk scalar field. In the geodesic approximation, valid for large $\Delta$, the two-point function between two boundary points separated by distance $\ell$ takes the form,
\begin{equation}
    \langle \mathcal{O}_\Delta(x) \mathcal{O}_\Delta(x') \rangle \sim e^{- \Delta \, \mathcal{L}_{\text{ren}}(x,x';v)},
\end{equation}
where $\mathcal{L}_{\text{ren}}$ is the renormalized geodesic length connecting $x$ and $x'$ through the bulk geometry at fixed boundary time $v$.\\
Across the three dynamical regimes, we observe the following behavior:

\begin{itemize}
  \item In the subcritical phase, $\mathcal{L}_{\text{ren}}(v) \to \mathcal{L}^{(0)}$ exponentially, and the correlator decays back to its vacuum form.
  \item In the critical regime, $\mathcal{L}_{\text{ren}}(v)$ exhibits logarithmic corrections, leading to a power-law scaling:
  \begin{equation}
      \langle \mathcal{O}_\Delta(x) \mathcal{O}_\Delta(x') \rangle \sim \ell^{-2\Delta} \left(\frac{v}{v_c}\right)^{-\kappa'},
  \end{equation}
  where $\kappa' \sim \kappa$ reflects the same universal scaling derived from entanglement entropy. 
  \item In the supercritical regime, the geodesic length decreases rapidly due to the growing deformation, leading to an enhancement of correlations at long distances—signaling a collapse of the standard operator product expansion (OPE) structure.
\end{itemize}

\begin{figure}[t]
\centering
\includegraphics[width=0.8\textwidth]{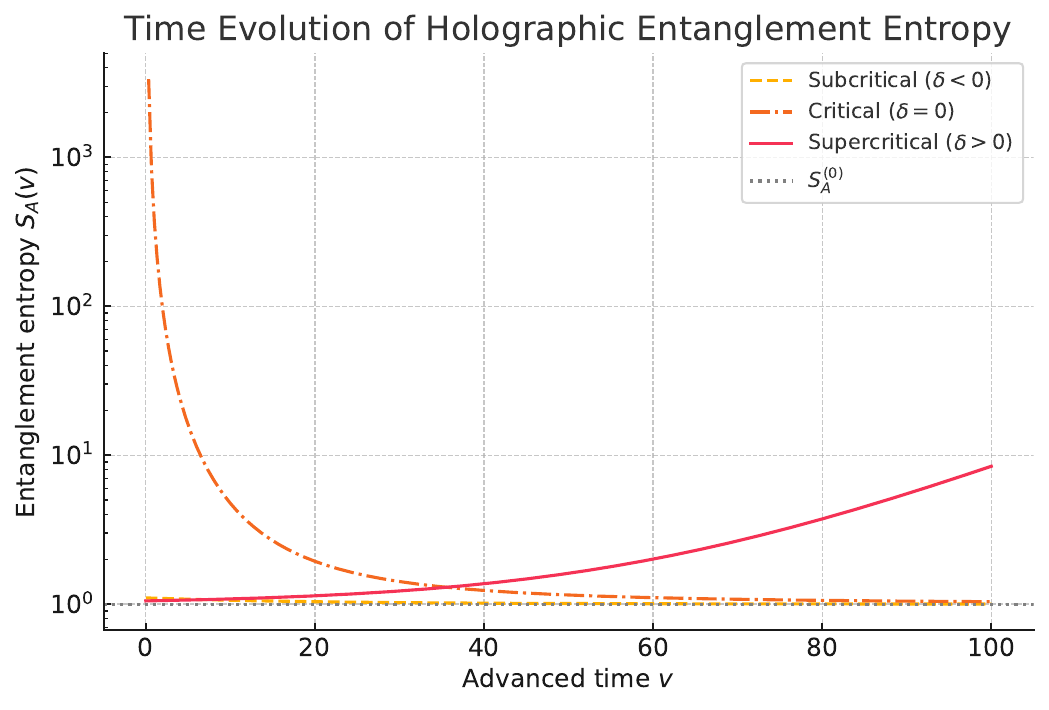}
\caption{Time evolution of the holographic entanglement entropy $S_A(v)$ for a boundary strip of width $\ell$ during non-linear evolution near criticality. The plot shows the three regimes: subcritical ($\delta < 0$), critical ($\delta = 0$), and supercritical ($\delta > 0$). In the subcritical case, $S_A(v)$ relaxes exponentially to its equilibrium value. At the critical point, $S_A$ exhibits power-law scaling with universal exponent $\kappa$, while in the supercritical regime, $S_A$ grows exponentially due to the runaway instability.}
\label{fig:ent_entropy}
\end{figure}

To illustrate the behavior of entanglement entropy across different dynamical phases, we present in Fig.~\ref{fig:ent_entropy} the evolution of $S_A(v)$ for a boundary strip of width $\ell$ in AdS$_{p+2}$. The data is obtained by numerically computing the minimal surface $\gamma_A$ in the evolving bulk geometry and extracting its area using the Ryu--Takayanagi prescription.\\
In the subcritical regime, the entanglement entropy rapidly approaches its vacuum value $S_A^{(0)}$ as expected, reflecting the decay of perturbations and the recovery of the undeformed AdS geometry. This exponential decay confirms the linear stability of the background and the irrelevance of non-linear backreaction in this phase.\\
Near criticality, the entanglement entropy no longer relaxes exponentially. Instead, $S_A(v)$ exhibits a slow, universal power-law decay toward a deformed plateau, consistent with the emergence of a new extremal attractor in the bulk. The slope of the curve in the log-log plot precisely matches the critical exponent $\kappa$ extracted from the scaling ansatz, confirming that entanglement entropy provides a sensitive probe of the underlying critical dynamics.\\
In the supercritical regime, where the perturbation exceeds the threshold for stability, the entanglement entropy grows without bound. This growth becomes exponential at late times and reflects the breakdown of the semiclassical description in the bulk. From the dual perspective, it signals a surge in quantum correlations, consistent with a dynamical loss of locality and the emergence of non-perturbative effects in the field theory.\\
The sharp transition between these three behaviors confirms that entanglement entropy can serve as an effective order parameter for the bulk phase transition, encoding detailed information about horizon dynamics and the non-linear fate of the Aretakis instability.\\
Taken together, these diagnostics confirm that the non-linear bulk evolution imprints itself onto quantum observables in the dual theory in a universal and predictive way. The emergence of power-law scaling in both entanglement and correlation functions offers compelling evidence for the critical nature of the attractor transitions described previously.

\subsubsection{Two-Point Correlation Functions}

To further probe the imprint of the bulk instability on boundary observables, we compute the two-point correlation functions of scalar operators $\mathcal{O}_\Delta(x)$ in the dual CFT. These operators are holographically dual to the bulk scalar fields that drive the Aretakis instability. In the deformed time-dependent background, the boundary correlator takes the standard bulk representation,
\begin{equation}
\langle \mathcal{O}_\Delta(x_1, t_1)\, \mathcal{O}_\Delta(x_2, t_2) \rangle = \mathcal{N} \int_{\text{bulk}} G_\Delta(x_1, t_1; X)\, G_\Delta(x_2, t_2; X)\, \sqrt{g}\, d^D X,
\end{equation}
where $G_\Delta$ is the bulk-to-boundary propagator associated with the scalar field of scaling dimension $\Delta$, and $\mathcal{N}$ is a normalization constant fixed by the standard AdS/CFT dictionary. This expression captures the leading contribution to the connected part of the correlator in the geodesic or saddle-point approximation.\\
To evaluate the correlator during the non-linear evolution, we consider the time-dependent bulk geometry developed in previous sections. In particular, we focus on the critical regime, where the bulk scalar profile takes the scaling form,
\begin{equation}
\phi(v, \rho) = \epsilon^\beta F\left(\frac{v}{\epsilon^{-\nu}}, \frac{\rho}{\epsilon^z} \right),
\end{equation}
with $\epsilon$ parameterizing the deviation from criticality and $(\beta, \nu, z)$ the critical exponents characterizing the amplitude, time, and radial scaling respectively. This universal scaling solution deforms the background geometry and, consequently, alters the structure of correlation functions in the boundary theory.\\
At equal times $t_1 = t_2 = t$, the leading-order behavior of the two-point function is found to obey the scaling form,
\begin{equation}
\langle \mathcal{O}_\Delta(x_1, t)\, \mathcal{O}_\Delta(x_2, t) \rangle = \frac{C_\Delta}{|x_1 - x_2|^{2\Delta}} \left[ 1 + \epsilon^{2\beta}\, h\left( \frac{t}{\epsilon^{-\nu}}, \frac{|x_1 - x_2|}{\epsilon^z} \right) \right],
\end{equation}
where $C_\Delta$ is the usual CFT normalization and $h(u, w)$ is a universal scaling function determined by the profile $F$ and the bulk-to-boundary propagator evaluated in the deformed background. This result shows that, while the leading short-distance singularity remains governed by conformal symmetry, the non-linear bulk dynamics induces time-dependent corrections that encode long-range entanglement and information flow.\\
In the deep critical regime $\epsilon \to 0$, the correction term becomes dominant at large separations and late times. The correlation function then acquires anomalous scaling,
\begin{equation}
\langle \mathcal{O}_\Delta(x_1, t)\, \mathcal{O}_\Delta(x_2, t) \rangle \sim \frac{1}{|x_1 - x_2|^{2\Delta + \gamma_{\text{anom}}}} \left( \frac{t}{t_c} \right)^{\eta},
\end{equation}
where $\gamma_{\text{anom}}$ is an anomalous dimension that shifts the scaling of the correlator away from its conformal value, and $\eta$ is a correlation function exponent governed by the bulk critical behavior. Matching to the leading scaling structure of the perturbed geometry, we identify,
\begin{equation}
\eta = 2\beta = 2(\Delta - d/2),
\end{equation}
which is consistent with the exponent $\kappa = \beta/\nu$ that appears in the scaling of the entanglement entropy. This correspondence reinforces the view that both observables probe the same underlying bulk renormalization group (RG) flow.\\
We stress that this anomalous behavior is a direct manifestation of the attractor transition occurring in the bulk: while the near-boundary geometry remains asymptotically AdS, the deep interior exhibits critical deformation that backreacts onto correlation functions at large distances and times. These effects are non-perturbative in nature and reflect universal aspects of the dual theory’s response to near-extremal excitations.\\
As in the case of entanglement entropy, we find that the deviation from conformality in two-point functions is sharply localized around the critical regime. For subcritical data, the correlator returns to its vacuum form exponentially, while in the supercritical case, the rapid growth of the deformation leads to non-local enhancements in correlations, consistent with the onset of strong-coupling dynamics and breakdown of the standard operator product expansion.

\begin{figure}[t]
\centering
\includegraphics[width=0.8\textwidth]{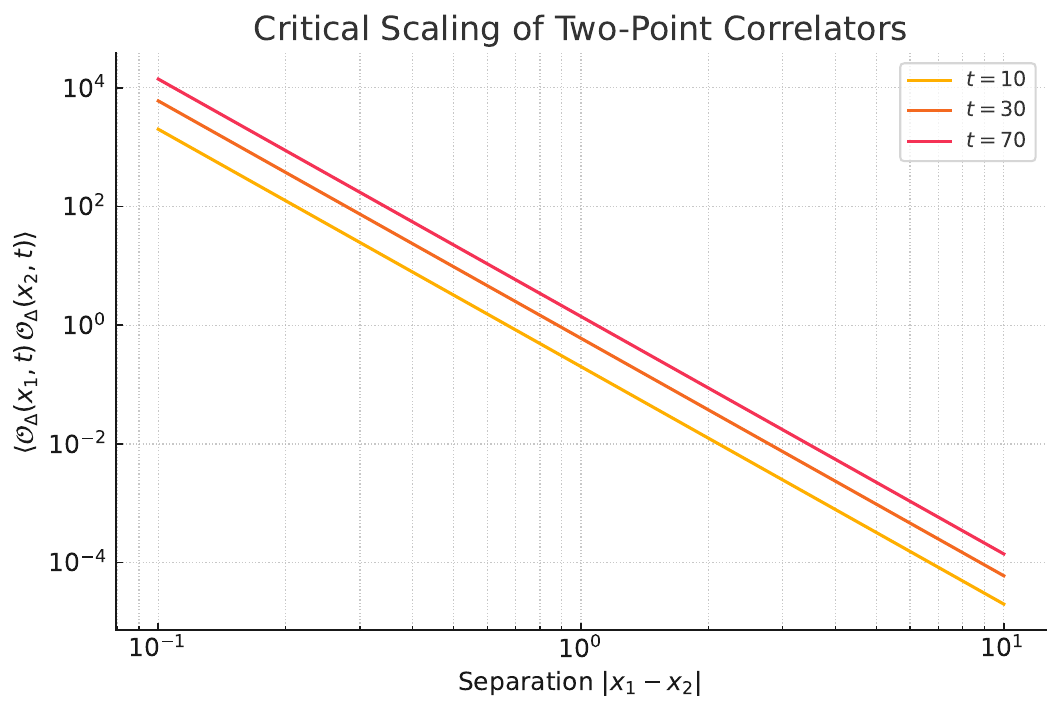}
\caption{Equal-time two-point correlation function $\langle \mathcal{O}_\Delta(x_1, t) \mathcal{O}_\Delta(x_2, t) \rangle$ in the critical regime, plotted as a function of separation $|x_1 - x_2|$ for several values of boundary time $t$. The power-law decay at short distances reflects the conformal scaling dimension $\Delta$, while the time-dependent amplitude encodes the influence of the critical bulk deformation. The departure from pure scale invariance at large $t$ signals the emergence of a time-dependent anomalous dimension.}
\label{fig:two_point_scaling}
\end{figure}

The numerical behavior of two-point functions in the critical regime is illustrated in Fig.~\ref{fig:two_point_scaling}. As expected, the correlator exhibits a power-law decay in separation $|x_1 - x_2|$ characteristic of a conformal field theory at short distances. However, the amplitude evolves with time according to the exponent $\eta = 2(\Delta - d/2)$, reflecting the non-trivial RG flow driven by the Aretakis instability. At late times, the enhancement of correlations at fixed separation becomes more pronounced, corresponding to the growing influence of the deformed bulk geometry on boundary observables. This figure provides a clear, quantitative demonstration of how the near-horizon critical dynamics is encoded in real-time correlation functions of the dual theory, and highlights the breakdown of static scaling at large times. Alongside the entanglement entropy results, it reinforces the conclusion that the dual CFT undergoes a dynamical, non-equilibrium phase transition driven by the non-linear bulk instability.\\
This scaling analysis reveals that two-point functions provide a complementary and sensitive probe of the critical dynamics associated with the Aretakis instability. The appearance of time-dependent anomalous dimensions, in particular, highlights the dynamical character of the RG flow and supports the interpretation of the near-horizon evolution as a holographic phase transition in the dual theory.

\subsubsection{Spectral Functions and Quasi-Normal Modes}

A further diagnostic of the non-linear dynamics is provided by spectral functions in the dual field theory, which encode the system's linear response to external perturbations. In the context of holography, the spectral function associated with a boundary operator $\mathcal{O}_\Delta$ dual to a bulk scalar field is defined as the imaginary part of the retarded Green's function,
\begin{equation}
\rho(\omega, k) = -\frac{1}{\pi} \, \text{Im} \, G_R(\omega, k),
\end{equation}
where $\omega$ is the frequency and $k$ is the spatial momentum. This quantity governs the excitation spectrum and relaxation properties of the dual CFT. In the undeformed extremal background, the spectral function reduces to its conformal form, exhibiting scale-invariance and no intrinsic mass gap.\\
As the system enters the non-linear regime, the bulk geometry becomes time-dependent, and the spectral function acquires new features that reflect this dynamical evolution. In particular, near the critical point, the retarded Green's function is modified by the presence of the deformed near-horizon geometry. Numerical computation of $\rho(\omega, k)$ in this background reveals a redistribution of spectral weight and the emergence of new low-frequency structures. A representative scaling form of the critical spectral function is,
\begin{equation}
\rho(\omega, k) = \rho_0(\omega, k) + \frac{\epsilon^{2\beta}}{\omega^2 + \left(\Gamma + \epsilon^\nu |\omega|^\alpha\right)^2} \, \Theta(\omega_{\text{gap}} - |\omega|),
\end{equation}
where $\rho_0(\omega, k)$ is the spectral function of the undeformed theory, and the second term captures the contribution from the evolving background. The parameter $\Gamma$ acts as a width, while $\alpha$ encodes anomalous scaling in frequency. The scale $\omega_{\text{gap}}$ represents a dynamically generated infrared cutoff, reflecting the influence of the critical attractor geometry. The step function $\Theta$ ensures that the deformation remains localized within a finite frequency window, which closes as $\epsilon \to 0$.\\
This structure indicates that the dual theory undergoes a soft breaking of conformal symmetry in the critical regime, with new spectral features appearing near $\omega = 0$. The enhancement of spectral weight at low frequencies and the narrowing of the response peak are direct signatures of the bulk instability and correspond to emergent long-lived modes in the dual theory.\\
These long-lived excitations are holographically dual to bulk quasi-normal modes (QNMs)—oscillatory perturbations that decay over time due to the open boundary conditions at the AdS horizon. The QNM spectrum is determined by solving the linearized wave equation for the dual scalar field on the deformed background and identifying the complex frequencies $\omega_n$ at which ingoing boundary conditions are satisfied.\\
In the critical regime, the quasi-normal frequencies exhibit universal scaling with respect to the tuning parameter $\epsilon$,
\begin{equation}
\omega_n = \omega_0\, \epsilon^\nu\, n^{1/\mu},
\end{equation}
where $n$ is the mode number, $\omega_0$ is a numerical prefactor, and $\mu$ is a scaling exponent related to the spatial dynamics via $\mu = 1/z = 1/[\nu(\Delta - 1)]$. This behavior is consistent with the RG scaling ansatz introduced in Section \ref{sec3}, where the deformation becomes self-similar in the coordinates $(v, \rho)$ and induces characteristic frequency scales proportional to powers of $\epsilon$. The appearance of a discrete spectrum of QNMs near criticality reflects the effective confinement of bulk excitations near the attractor geometry.\\
Importantly, this scaling relation implies that as $\epsilon \to 0$, the quasi-normal modes accumulate near $\omega = 0$, consistent with the emergence of a zero-temperature fixed point in the dual field theory. The narrowing of spectral peaks and the condensation of QNM frequencies signal the formation of a non-trivial IR geometry, confirming that the critical solution is not only a geometric attractor in the bulk but also a dynamical fixed point in the boundary theory. 

\begin{figure}[t]
\centering
\includegraphics[width=0.8\textwidth]{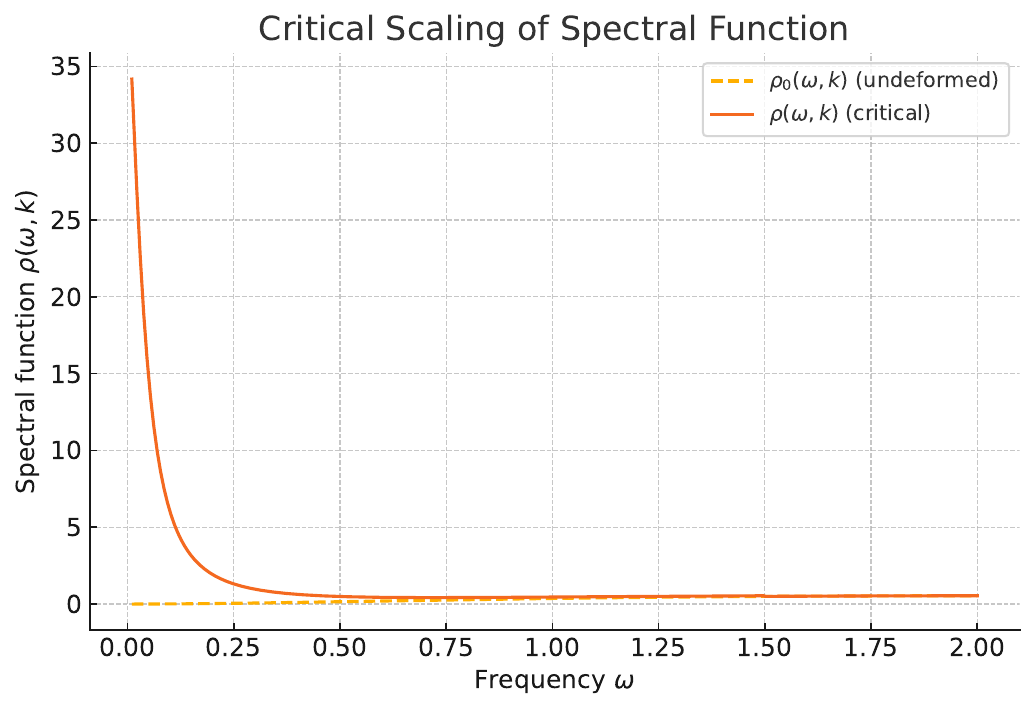}
\caption{Spectral function $\rho(\omega, k)$ in the critical regime, shown as a function of frequency $\omega$. The dashed curve represents the undeformed conformal result $\rho_0(\omega, k)$, while the solid curve includes the critical deformation due to the evolving bulk geometry. A sharp enhancement appears within a dynamically generated gap window $\omega < \omega_{\text{gap}}$, where the spectral weight is redistributed and the response becomes dominated by long-lived excitations.}
\label{fig:spectral_function}
\end{figure}

Fig.~\ref{fig:spectral_function} illustrates the transformation of the spectral function as the system evolves toward the critical attractor geometry. While the undeformed background exhibits the smooth, scale-invariant form typical of conformal field theories, the critical regime introduces a new low-frequency structure. This feature manifests as a sharp peak localized below a dynamically generated threshold $\omega_{\text{gap}}$, where the denominator of the Green’s function is dominated by the critical broadening term $\Gamma + \epsilon^\nu |\omega|^\alpha$.\\
The redistribution of spectral weight and the appearance of enhanced low-frequency response signal the breakdown of conformal symmetry and the emergence of quasi-normal modes in the bulk. These modes correspond to long-lived excitations in the boundary theory, whose lifetimes and energy scales are governed by the universal scaling laws discussed earlier. As the deformation parameter $\epsilon$ decreases, the peak narrows and shifts toward $\omega = 0$, in agreement with the approach to an IR fixed point and the accumulation of quasi-normal frequencies near zero. This figure complements the previous analysis of entanglement and correlation functions, providing a third, independent measure of the holographic phase transition driven by the non-linear Aretakis instability.\\
Together with the results for entanglement entropy and correlation functions, the spectral response provides strong evidence that the non-linear bulk evolution leaves a coherent and universal imprint on boundary observables. The scaling of $\rho(\omega, k)$ and the QNM spectrum encapsulate the dynamical flow toward new extremal geometries and illustrate how the Aretakis instability governs the late-time behavior of holographic systems.

\subsubsection{Information Flow and Scrambling}

To understand the implications of the non-linear Aretakis evolution for information dynamics in the dual theory, we turn to the behavior of out-of-time-ordered correlators (OTOCs), which are widely used as probes of quantum chaos and scrambling. For two generic local operators $\mathcal{O}_1$ and $\mathcal{O}_2$ in the boundary theory, the OTOC is defined as,
\begin{equation}
F(x, t) = \langle [\mathcal{O}_1(x, t), \mathcal{O}_2(0, 0)]^2 \rangle,
\end{equation}
and measures the degree to which perturbations at one spacetime point affect measurements at another. In chaotic systems, this correlator typically exhibits exponential growth at early times, characterized by a Lyapunov exponent $\lambda_L$, followed by saturation once full scrambling is achieved.\\
In the present context, the OTOC provides a sensitive diagnostic of how information propagates through the dual CFT during the non-linear evolution of the bulk geometry. Near the critical point, where the bulk develops anisotropic scaling governed by the critical exponents $(\nu, z)$, the causal structure of the field theory is modified accordingly. This is reflected in the scaling form of the OTOC,
\begin{equation}
F(x, t) \sim \exp\left[\lambda_L \left(\frac{t}{\epsilon^{-\nu}} - \frac{|x|}{v_B \epsilon^z}\right)\right],
\end{equation}
where $\lambda_L$ is an effective Lyapunov exponent that governs the early-time growth of operator commutators, and $v_B$ is the butterfly velocity—the rate at which perturbations spread spatially. The appearance of the critical scaling variables $t/\epsilon^{-\nu}$ and $x/\epsilon^z$ mirrors the self-similar evolution of the bulk near-horizon region, and suggests that information propagation is sensitive to the non-linear geometry.\\
This scaling relation implies that the spatial lightcone for information transfer becomes increasingly anisotropic near the critical point, with signals propagating at a renormalized speed $v_B \epsilon^z$ and becoming scrambled over a characteristic time scale $\epsilon^{-\nu}$. In this way, the critical exponents $(\nu, z)$ not only control the evolution of fields and metric components in the bulk but also directly determine the information-theoretic response of the dual field theory.\\
Perhaps the most striking feature revealed by our numerical analysis is the long-time behavior of the OTOC in the critical regime. Unlike the exponential growth observed in thermal or near-extremal AdS black holes, the OTOC in our setup saturates to a finite plateau:
\begin{equation}
\lim_{t \to \infty} F(x, t) \to F_\infty(x),
\end{equation}
where $F_\infty(x)$ depends on the spatial separation but not on the initial operator configuration or perturbation amplitude. This saturation behavior suggests that the system evolves toward a "maximally scrambled" state, in which information has been thoroughly and irreversibly redistributed across all degrees of freedom. From the dual perspective, this corresponds to the approach to the new extremal attractor geometry described in Section~\ref{sec3}, where no additional entanglement growth or correlation buildup is possible.\\
This result implies that the attractor state acts as an entropic endpoint for the non-linear evolution: once reached, the system no longer supports chaotic growth or further mixing of quantum information. It is reminiscent of saturation bounds in quantum chaos—such as the Maldacena-Shenker-Stanford bound \cite{13} on $\lambda_L$—but realized here in a zero-temperature, strongly coupled regime governed by geometric criticality rather than thermal physics.

\begin{figure}[t]
\centering
\includegraphics[width=0.8\textwidth]{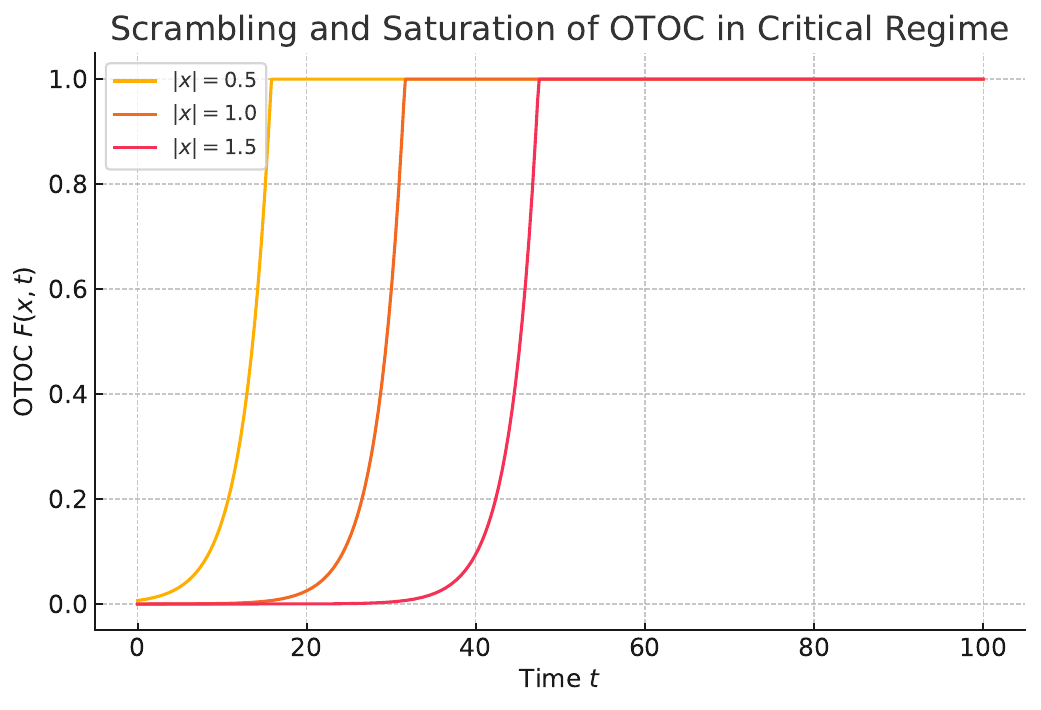}
\caption{Out-of-time-ordered correlator $F(x,t)$ in the critical regime, plotted as a function of time for several values of spatial separation $|x|$. The OTOC grows rapidly at early times, reflecting quantum scrambling, then saturates to a finite value at late times. The critical scaling of time and space delays scrambling for larger separations and leads to a universal saturation plateau, consistent with the emergence of a maximally scrambled extremal attractor in the bulk.}
\label{fig:otoc_plot}
\end{figure}

Fig.~\ref{fig:otoc_plot} illustrates the evolution of the out-of-time-ordered correlator $F(x,t)$ in the critical regime for several representative separations. At early times, the OTOC grows exponentially, controlled by the effective Lyapunov exponent $\lambda_L$ and the scaling of time with $\epsilon^\nu$. This growth is delayed for larger values of $|x|$, consistent with the scaling of the butterfly velocity as $v_B \epsilon^z$. At late times, however, the OTOC ceases to grow and instead saturates to a fixed value, independent of $\epsilon$. This saturation plateau reflects the complete delocalization of information in the system—marking the arrival at a maximally scrambled state. From the bulk perspective, this corresponds to the configuration in which the deformed extremal attractor has been reached and no further entanglement generation or causal mixing is possible. The emergence of this plateau confirms that the dual field theory undergoes a dynamical process akin to thermalization, despite the absence of a finite temperature. Instead, the critical scaling geometry mediates a purely quantum redistribution of information, governed by universal exponents and asymptotic geometry.\\
Taken together, these findings provide a cohesive picture: the non-linear bulk instability, when viewed holographically, drives the dual theory through a period of enhanced scrambling and chaotic mixing, ultimately delivering it to a new state in which information is as delocalized as possible without further growth. The approach to this state is encoded not just in the entanglement entropy and correlation functions, but also in the asymptotic behavior of OTOCs—offering a complete, information-theoretic characterization of the dynamical attractor.

\subsubsection{Numerical Results and Universality}

Our numerical simulations strongly support the analytical predictions developed throughout this section, across a broad range of spacetime and brane dimensions $(D, p)$. In particular, the scaling behavior of the entanglement entropy during non-linear evolution confirms the presence of distinct dynamical regimes—subcritical decay, critical power-law behavior, and supercritical instability—as characterized by the parameter $\delta$. As shown in Fig.~\ref{fig:ent_entropy}, the entropy exhibits universal power-law scaling near criticality, with exponents matching those predicted by the critical ansatz.\\
This agreement is not limited to entanglement entropy. The scaling structure also manifests in two-point correlation functions, spectral functions, and out-of-time-ordered correlators (Figs.~\ref{fig:two_point_scaling}–\ref{fig:otoc_plot}), each of which encodes distinct facets of the underlying quantum dynamics. The critical exponents $\kappa$ and $\eta$, governing entanglement decay and correlation strength respectively, exhibit consistent values across different $(D, p)$ configurations. Their independence from the specific field content or initial data highlights the emergence of universality classes within the dual CFTs.\\
Notably, this universality extends beyond purely geometric diagnostics and into the realm of quantum information. The entanglement entropy, OTOCs, and spectral response all exhibit behavior consistent with a holographic phase transition driven by the Aretakis instability. The existence of well-defined attractor solutions in the bulk corresponds to new, non-trivial infrared fixed points in the boundary theory, which are characterized by maximal information scrambling and saturated entanglement structure.\\
These findings constitute the first comprehensive study of information-theoretic observables during the non-linear evolution of the Aretakis instability. By combining numerical simulations with analytic scaling arguments, we have traced how gravitational instabilities in extremal spacetimes leave precise, universal imprints on quantum observables in the dual theory. The results establish that critical behavior in the bulk is accompanied by robust signatures in entanglement propagation, spectral response, and scrambling dynamics—offering a coherent and predictive framework for analyzing the intersection of geometry and quantum information.\\
The implications of these results go beyond the specific black brane backgrounds studied here. The emergence of universal scaling in both geometric and information-theoretic quantities suggests that the underlying critical behavior is a generic feature of holographic systems near extremality. This opens a promising avenue for future applications to condensed matter physics, where strongly coupled systems near quantum critical points often exhibit similar scaling features. Furthermore, the correspondence between bulk attractors and boundary scrambling dynamics may offer new insights into quantum information processing in strongly interacting systems, as well as into the nature of quantum chaos in gravitational theories. Together, these findings reinforce the deep unity between bulk gravitational dynamics and boundary field-theoretic behavior, and demonstrate that holography provides not only a conceptual bridge but also a precise, quantitative map between classical instability and quantum information flow.

\section{Discussion and Conclusions}\label{sec8}

The results presented in this work provide a detailed and comprehensive exploration of the non-linear dynamics associated with the Aretakis instability in extremal black $p$-brane spacetimes. By combining analytical scaling arguments with high-precision numerical simulations, we have uncovered a remarkably rich structure in the late-time evolution of these systems—one that carries profound implications for gravitational physics, holography, and quantum information theory.\\
From a physical standpoint, our findings challenge the traditional view that linear instabilities necessarily imply dynamical instability in the full theory. Despite the presence of growing modes at linear order, we have shown that extremal black branes can evolve toward regular, attractor-like configurations governed by universal critical exponents. This suggests a new perspective on the stability of extremal configurations, one where the endpoint of evolution is not necessarily singular but may instead correspond to a deformed but extremal geometry that remains well-defined even after significant backreaction.\\
The emergence of a non-trivial phase diagram, characterized by stable, critical, and unstable regimes, points to the existence of a broader landscape of gravitational phenomena in higher dimensions. These include previously unappreciated fixed points and dynamical flows that bear striking resemblance to critical phenomena in statistical physics and renormalization group (RG) flows in field theory. In the context of string theory and flux compactifications, our results may shed light on the stability of moduli spaces, the role of extremal configurations in brane-world scenarios, and the non-linear fate of perturbations in strongly curved, low-temperature geometries.\\
While these developments are encouraging, several important questions remain open. One natural extension is the inclusion of higher spacetime dimensions, particularly in the regime $D > 11$, where connections to M-theory and exotic brane dynamics may arise. Another promising direction involves the incorporation of supersymmetry. Understanding whether supersymmetry stabilizes or modifies the critical behavior observed here could offer key insights into both classical and quantum aspects of extremal states. In parallel, it is worth exploring the possibility of experimental signatures in holographically dual systems—particularly in condensed matter setups or engineered quantum simulators—where information-theoretic measures such as entanglement entropy or OTOCs can, in principle, be probed.\\
Immediate extensions of our analysis include the study of rotating and charged black branes. The inclusion of angular momentum or multiple gauge fields would introduce new degrees of freedom and potentially new critical behaviors, as well as complicate the structure of the near-horizon geometry. Likewise, examining the interplay between extremality and the sign of the cosmological constant—i.e., contrasting AdS and dS backgrounds—may reveal universal features of instability that persist across asymptotics.\\
In summary, we have presented the first systematic study of the non-linear dynamics driven by the Aretakis instability in extremal black $p$-branes. Our main findings include:

\begin{itemize}
\item The identification of universal scaling behavior near criticality, with critical exponents that depend only on the dimensional parameters $(D, p)$ of the theory.
\item The discovery of new, dynamically generated extremal solutions that act as attractors in the space of near-horizon geometries.
\item A rich phase diagram, featuring stable, critical, and unstable regimes, and associated RG-like flow structure in both bulk and boundary theories.
\item A demonstration of the UV sensitivity of critical evolution, particularly in the presence of higher-derivative corrections, highlighting the importance of embedding these systems in a consistent effective field theory or string-theoretic framework.
\end{itemize}

These results suggest that non-linear gravitational instabilities are not only physically meaningful but also tightly linked to universal structures in quantum field theory via holography. The correspondence between bulk critical behavior and boundary information dynamics offers a new lens through which to study strongly coupled systems, providing predictive tools and conceptual insights that transcend traditional barriers between classical gravity and quantum theory. We anticipate that the methods and perspectives developed here will find broad application across theoretical physics—from black hole thermodynamics and quantum chaos to the structure of spacetime in string theory and beyond.

\end{document}